# Modification of Question Writing Style Influences Content Popularity in a Social Q&A System

**Igor A. Podgorny**
Intuit, Inc.
San Diego, USA
igor_podgorny@intuit.com

**ABSTRACT**
TurboTax AnswerXchange is a social Q&A system supporting users working on federal and state tax returns. Using 2015 data, we demonstrate that content popularity (or number of views per AnswerXchange question) can be predicted with reasonable accuracy based on attributes of the question alone. We also employ probabilistic topic analysis and uplift modeling to identify question features with the highest impact on popularity. We demonstrate that content popularity is driven by behavioral attributes of AnswerXchange users and depends on complex interactions between search ranking algorithms, psycholinguistic factors and question writing style. Our findings provide a rationale for employing popularity predictions to guide the users into formulating better questions and editing the existing ones. For example, starting question title with a question word or adding details to the question increase number of views per question. Similar approach can be applied to promoting AnswerXchange content indexed by Google to drive organic traffic to TurboTax.

**ACM Classification Keywords**
H.3.3. Information Search and Retrieval: Clustering.

**INTRODUCTION**
The ability to generate good quality content for an unlimited number of topics explains the increasing popularity of social question-and-answer (Q&A) systems. Given the complexity of the US tax code, social Q&A systems provide a sensible self-support option for tax and financial software where long-tail content generated by the users can supplement conventional frequently asked questions (FAQs). AnswerXchange (http://ttlc.intuit.com) is a social Q&A site where existing and prospective customers can learn and share their knowledge with other TurboTax (http://turbotax.intuit.com) users while preparing their federal and state tax returns. As the users step through the TurboTax interview screens, they can ask questions about software and tax topics (Figure 1) and receive answers in a matter of minutes. AnswerXchange was launched in 2007 under the name of TurboTax Live Community and has generated millions of questions and answers that helped tens of millions of TurboTax customers.

AnswerXchange users may also contribute to the site by commenting on questions and answers, and by voting answers up or down. The majority of views and votes come not from the users who ask the questions, but from the users who find the answers by search. In good agreement with the "1% rule", the askers represent only about 1% of AnswerXchange users, while 99% of users find answers by searching (Figure 2) and navigating AnswerXchange content. The AnswerXchange search engine is built with the Apache Lucene open-source search software [14]. Answered, commented or unanswered questions are all referred to below as the AnswerXchange questions.

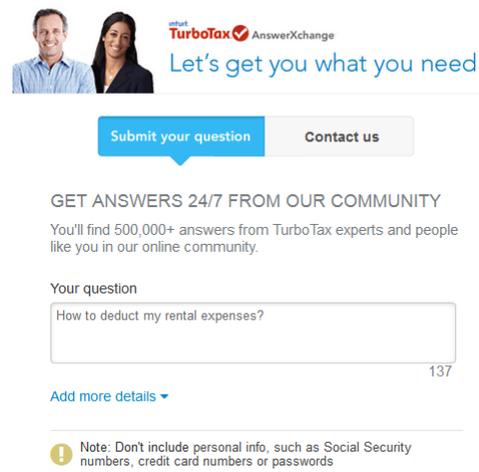

**Figure 1. AnswerXchange question-asking experience. Question summary is mandatory and limited to 170 characters. Question details are optional and unlimited in size.**

Understandably, the AnswerXchange operations are aligned with TurboTax business goals which differentiates AnswerXchange from conventional social Q&A systems such as Quora or Yahoo Answers. TurboTax Online is the "try-before-you-buy" software and so the first task of AnswerXchange is to help increase product conversion yet to drive down the assisted support costs. This is achieved by providing both high quality FAQs and user generated content. The second business goal is to engage existing TurboTax customers by increasing overall AnswerXchange page views and also to bring new customers to the AnswerXchange by leveraging organic traffic from Google and other search engines.

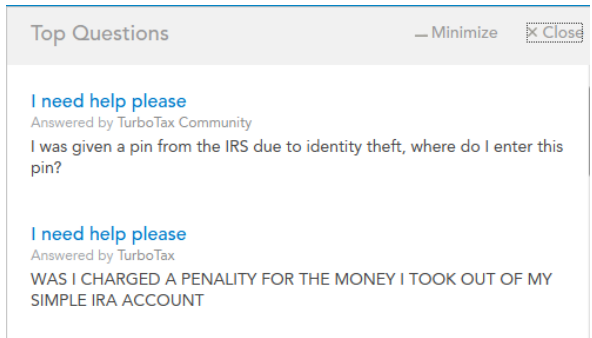

**Figure 2. AnswerXchange search results rendered in response to the user query "I need help". Question summaries and details are shown in blue and black, respectively.**

AnswerXchange content is a combination of tax and product related topics. Tax questions are semantically similar to publications by the Internal Revenue Service and state tax authorities. Product questions are TurboTax specific and deal with pricing, choice of product version, and software issues such as installation or e-filing. The content type determines how the question is routed to answering. Answers to the tax questions come from non-compensated "super users" selected by moderators based on frequency and quality of contributions. The product related questions are typically answered by support agents who are salaried employees and are trained to write good quality answers. User vote statistics is one of the metrics tracked by the AnswerXchange and so the predictability of user votes in the social Q&A systems can be leveraged by the AnswerXchange operations to improve quality of answers.

On the contrary, the AnswerXchange questions are written by regular users who often lack domain knowledge and proper writing skills. Yet this is the question that is shown to the searching users before they can even see the answer (Figure 2). Question features may influence user's intent to click on the question snippet and hence content popularity defined as number of views accumulated by the question during the tax season. The AnswerXchange content popularity dynamics is complex and driven by competition for TurboTax customers' attention. Specifically, AnswerXchange content competes with curated FAQs and assisted support options when TurboTax customers search for general topics and with long-tail tax related content indexed by Google outside of TurboTax. Social factors also play some role as super users and agents may be selective in deciding which question to answer trying to maximize the number of up votes from the askers.

## RESEARCH QUESTIONS

Predicting popularity of online social content received a lot of interest in recent years [2, 22, 23] including the task of predicting content popularity before publication [23]. One common observation from online data analysis is that user attention largely follows power law probability distribution, with only a small fraction of content receiving the most users' attention. Answer [1, 9, 13, 20] quality and question utility in the social Q&A systems has been the focus of increasing attention as well. The research approaches included statistical analysis, natural language processing, expert ratings of the content quality, predictive modeling and also automated classification of user votes.

Good question typically results in a good answer, and at least one answer is typically required for the content to be exposed to search in order to become popular. Probability of question to be answered can be predicted as well [4] based on the model attributes similar to those used for predicting question quality.

AnswerXchange is a social Q&A platform with content popularity driven by variety of factors such as topics (or content categories), content quality and, to some degree, by question and answer writing style. The research goal of this paper is to explore relationship between question quality, utility and popularity based on AnswerXchange user votes and traffic data. We give the center stage to writing style of the question since it can be altered during question formulation process. The second goal of this paper is to identify question attributes that can be edited by the users to modify expected question popularity as part of the AnswerXchange operations.

## DATA AND ANALYSIS

### Data Set and Seasonality Factors

Users have three options to view the AnswerXchange content. Authenticated users typically consume content when it is shown in response to the user search queries inside TurboTax Online. Non-authenticated users working with TurboTax desktop version typically search for content shown at the AnswerXchange landing page (http://ttlc.intuit.com). The remaining option is to view the AnswerXchange content indexed by Google and other search engines. View and vote statistics are recorded by all three channels and subsequently used by the scoring algorithms when boosting content served by the AnswerXchange search engine or as part of the Google search engine optimization. For example, TurboTax users who are dissatisfied with the TurboTax prices may vote the related AnswerXchange content down. This, in turn, may reduce the Google views of this particular AnswerXchange question. In short, user search, user votes and click patterns inside TurboTax affect click patterns at AnswerXchange content indexed by Google and vice versa.

In this study, we have employed 289,000 questions created from 01/01 (when the user traffic starts to ramp up) to 04/15/2015 (Tax Day in 2015). 195,500 questions received at least one answer (67.5% answer rate) and 231,000 received at least one view. 45% of views came from the top 1% and 76% of views came from the top 10% of the questions. The bottom 20% of questions did not get a single view. The average number of views per question was 23.7 (32.1 for the answered ones) and any question with more than 34 views would make it to the top 10%. User activity in AnswerXchange exhibits strong seasonal patterns (Figure 3A) with the most popular content being generated during the first few weeks for the ax season (Figure 3B).

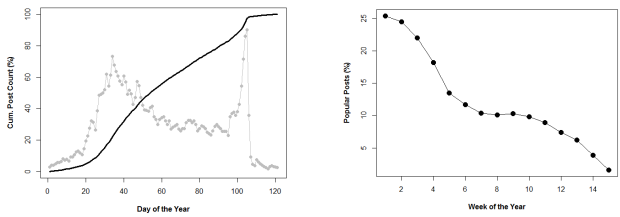

**Figure 3. Daily (gray line) and cumulative (black line) question counts (A). Fraction of questions in top 10% (B).**

Google organic traffic contributed 20.5% to all views on AnswerXchange content created during 2015 tax season, but more than 86% of the Google traffic came from the top 10% of the questions. We will hereafter refer to this set of data as the 2015 AnswerXchange dataset.

**Probabilistic Topic Model**

Probabilistic topic models have become widely accepted in recent years as a mechanism to discover hidden topics in the text and organize collection of the documents by the discovered themes [3]. While probabilistic topic models do not provide labels for the identified topics, the labels can be added by the experts with relatively little effort. Topics provide a convenient method for dimensionality reduction in the (unstructured) text data. Detecting the semantic coherence of a document is another task relevant to text segmentation and categorization and can be addressed based on the concept of topic entropy [12].

Tax and product content categories can be thought as two super topics, where the probability of a particular question belonging to either topic is a one-dimensional variable [15]. One-dimensional content model was shown to perform reasonably well in capturing variability of askers' votes in AnswerXchange [15, 16]. Increasing the number of topics is equivalent to substituting the one-dimensional content model by the multi-dimensional one. Towards this goal, we have applied Latent Dirichlet Allocation model [3] to identify 30 topics in the 2015 AnswerXchange dataset using R package "topic models" [6].

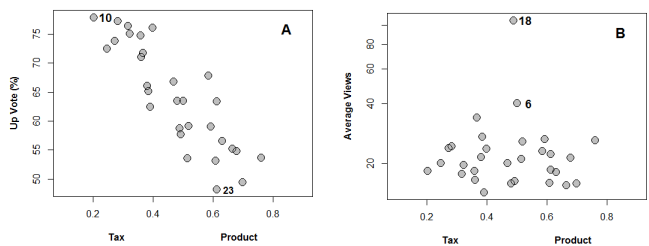

**Figure 4. Average up vote per topic (A) and average views per topic (B) vs. content type variable. Topics marked with numbers are explained in the text.**

Shown in Figure 4A are the askers' vote statistics per topic as a function of one-dimensional content variable computed based on the approach described in [15]. Up vote fraction is the ratio of up votes and all votes. Content class labels 0 and 1 mean that the question is tax or product related with 100% certainty, respectively. The strong negative correlation (Pearson's r=–0.866) between up vote fraction and content type variable validates one-dimensional approximation used in [15]. AnswerXchange users tend to down vote answers to the product related question more frequently as compared to answers to the tax related questions, often irrespective of the quality of answers. Content type was found to be the most important attribute, reflecting the difference in voting statistics between the tax and product related questions [15]. Topics 10 and 23 shown in Figure 4A illustrate this point further. Topic 10 is about claiming dependents (Table 1) and hence a well-defined tax topic. On the contrary, topic 23 about contacting assisted support, i.e. a well-defined product related topic.

| Topic | Top keywords | Summary examples |
|---|---|---|
| 6 | refund, account, check, receive, bank | I haven't received my refund. |
| 10 | claim, child, son, dependent, daughter | Can we claim my son who lived with us most of the year? |
| 18 | file, still, day, accept, say | How long will it take for the IRS to accept my taxes? |
| 23 | help, number, please, card, call | Want to talk to someone. |

**Table 1. Examples of AnswerXchange topics.**

Shown in Figure 4B are average numbers of question views versus one-dimensional content variable. By comparing Figures A and B one can see that high fraction of up votes in the topic (proxy metrics of content quality) does not result in an increased number of views (proxy metrics of content popularity) and vice versa. In fact, the correlation between askers' votes and question views is close to zero. This makes the task of judging the popularity specific to a topic, but not to tax versus product type. Topic 18 clearly stays apart from other topics shown in Figure 4B. Once it is excluded, correlation between average views and number of questions per topic (not shown) becomes close to zero. Topics 6 and 18 shown in Figure 4B belong to neither tax or product category and can be asked both inside and outside of TurboTax (Table 1). AnswerXchange questions belonging to topic 18 bring the largest share of Google organic traffic in both absolute and relative terms. In fact, for the questions belonging to topic 18, fraction of views attributable to Google search is 49.5% exceeding the average by 150%. The second most popular topic 6 also has an above average fraction of Google views of 29%. Topics 6 and 18 somewhat similar semantically, but topic 6 is more about questions on the expected refunds for the tax returns already accepted by IRS. Popularity of questions belonging to topics 18 and 6 is hardly surprising. Being in the middle between tax and product categories, they are still applicable to both TurboTax customers and other tax filers. On the other hand, tax specific content, while being potentially useful for the information seekers outside of TurboTax, may be found from variety of other sources by searching Google instead of AnswerXchange.

**Question Writing Style**

Out of 289,000 questions in the 2015 AnswerXchange dataset, 50% have details and this fraction goes up to 68% in the top 10% of the most popular ones. When details are added, average details length is 218 characters going up to 271 in the top 10%. The effect of question summary is just the opposite. When details are missing, average summary length is 100 characters, but the summary length goes down to 87 in the top 10% of the most popular questions. The next question to ask is how summary and details should be written in order to maximize question popularity? This problem is further illustrated in Figure 5 where numbers of questions and views are plotted against question length in character (the sum of question summary length and question details length). There is a clearly defined local maximum in number of questions around 170 characters (Figure 5A). Recall that this is the limit on number of characters for the question summary. Users may type text of the question up to the limit or cut and paste a question exceeding the limit from somewhere and then post it without adding details. The second maximum around 250 characters is for questions with details. When question counts are replace with view counts, the first maximum reduces in size, while the second maximum shifts towards 400 characters (Figure 5B).

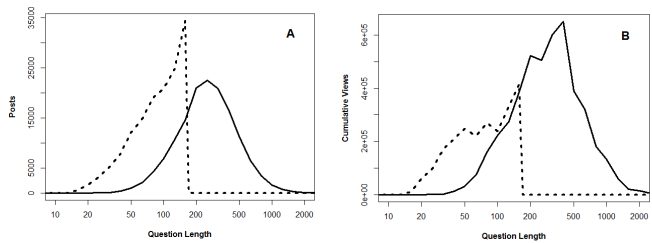

Figure 5. Number of questions (A) and cumulative count of views (B) vs. question length in characters. Statistics for questions with and without details are shown by solid and dashed lines, respectively.

This behavior is further illustrated in Figure 6A. When details are missing, shorter questions receive substantially higher number of views, but when details are added, longer questions typically are the winners. Shown in Figure 6B is coherency metrics (or topic entropy measure [12]). Topic entropy of question j is defined as the entropy of the posterior topic distribution [5]:

$$E_j = -\frac{1}{\log(M)} \sum_{k=1}^{M} \log(P_{jk}) P_{jk} \quad (1)$$

where $P_{jk}$ is probability of question j to belong to topic k and M is number of topics.

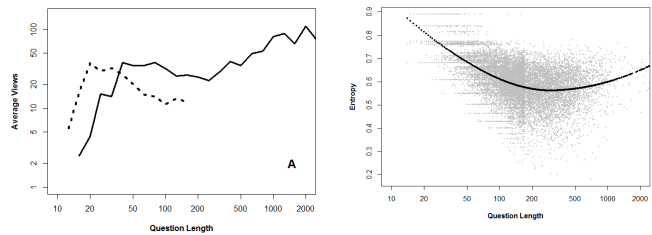

Figure 6. Average number of question views (A) vs. length for the questions with (solid line) and without (dashed line) details. Coherency metrics per question (gray dots) and average coherency metrics (B) vs. question length.

To explain the inverse relationship between summary length and popularity, recall how the Lucene ranking algorithm works [14]. In short, search results are more relevant if search query keywords are more frequent in the document (a concept known as "tf-idf" weighting and explained in text books about information retrieval). In such a way, the shorter summary with the same set of matching keywords always wins over the longer summary with the same semantic meaning. A typical AnswerXchange search query is about 36 characters (or 6.3 words) long, i.e. 50% longer than a typical Google query, but still shorter than a typical AnswerXchange question summary. The effect of question details can be explained the same way: questions with details receive an extra boost from Lucene and are therefore more likely to be shown at the top of the search results. The content shown in question details seems to play a minor role and the attentions patterns in online forums have been shown to focus mostly on question titles [25].

We now investigate the relationship between question summary features and question popularity. Shown in Table 2 are 20 most frequent first words of the summary for the questions from the 2015 AnswerXchange dataset. Before computing percentages, we (1) split "I'm" and "I've", (2) merged "Turbo" and "Tax" into a single term when needed, and (3) down cased all characters of the summaries. The percentages reported in Table 2 account for 76% of the questions. The most frequent token is "I" that is present in 27% of the questions, followed by the "knowledge" question words "how" and "why". The question popularity is measured by both the average number of views and by the probability for the question to be in the top 10% of the most popular questions. Both quantities are well correlated (Pearson's r=0.860, Spearman's ρ=0.846) and the questions starting with "does" have the highest average number of views for any group. Popularity (Top 10% column) and answer rate shown in Table 2 are also correlated (Pearson's r=0.616, Spearman's ρ=0.544). The answer rate dependence on the first word of the summary may reflect contributors' bias in selecting question for answering [17], but the answer rate does not vary with the same magnitude as popularity.

As seen from Table 2, the questions with summaries starting with "why" are almost four times less popular than the average pos. This is positive news from the AnswerXchange prospective since answers to the "why" questions are more frequently voted down [17]. With the exception of "why" and "when", which often indicate a rhetorical question type, the summaries starting with interrogative pronouns "where", "what" and "how" ("knowledge" questions [18]) and auxiliary verbs "are", "does", "can", "is" ("closed-ended" questions [18]) are more likely than average to become popular. Conversely, "declarative" summaries starting with "I", "my" and "we" are less likely than average to become popular. Closed-ended questions (e.g. "Can I deduct …?") are associated more often with tax related content and, if answered, receive a higher fraction of up votes. The "knowledge" question (e.g. "how" and "why") are more likely to be from the product category and, if answered, are more likely to be voted down. All words shown in Table 2 are legitimate words for starting grammatically correct English sentence (e.g. there are no coordinating conjunction "for", "and", "nor", "but", "or", "yet" and "so" in the list).

| First Word | Percentage | Views | Top 10% | Answer Rate |
|---|---|---|---|---|
| are | 0.4 | 29.1 | 17.0 | 72.1 |
| does | 0.7 | 40.9 | 16.2 | 74.5 |
| where | 3.5 | 37.6 | 15.6 | 73.4 |
| is | 1.5 | 23.4 | 14.1 | 71.5 |
| how | **10.8** | 30.0 | 14.0 | 74.3 |
| TurboTax | 1.2 | 28.4 | 13.5 | 65.3 |
| what | 3.8 | 38.1 | 13.3 | 68.3 |
| can | 4.1 | 23.7 | 12.6 | 80.5 |
| do | 1.8 | 26.1 | 12.1 | 76.6 |
| need | 0.7 | 23.6 | 11.0 | 68.4 |
| when | 1.4 | 31.1 | 10.1 | 72.0 |
| on | 0.6 | 19.0 | 8.8 | 57.6 |
| my | 5.6 | 23.9 | 8.1 | 71.5 |
| if | 1.9 | 18.0 | 7.5 | 77.3 |
| the | 1.1 | 22.7 | 7.3 | 60.6 |
| I | **27.0** | 15.6 | 7.0 | 69.1 |
| it | 0.6 | 11.1 | 6.2 | 58.5 |
| in | 0.6 | 11.5 | 6.1 | 61.7 |
| we | 1.0 | 15.0 | 6.0 | 69.8 |
| why | **8.0** | 8.6 | 2.3 | 51.4 |

Table 2. Percentages of questions (%), average views, percentages of questions in the top decile by popularity (%) and answer rate (%) vs. first word of the question summary.

In Table 2, 19 out of 20 words are common stop words [10]. The only exception is "TurboTax" – understandably one of the most frequent keywords in the AnswerXchange questions. Note that stop words (as any frequent terms in the document) are often discarded by the Lucene search engine. A typical AnswerXchange user spends 2-3 seconds on reading a search result snippet mostly focusing on the left top corner (a web marketing concept known as "Google golden triangle"). In this way, the user impression of the first few words of the summary (Figure 2) may play an important role in the user's intent to read the entire question. One explanation of the results shown in Table 2 is that AnswerXchange users are less interested in the content specific to somebody's experience (as revealed by the first person pronouns) and tend to skip the related search results. Psycholinguistic studies also confirm that human attentional focus largely depends on the pronoun usage [24].

The results presented in Table 2 do not necessarily mean that the first words of the summary are independent attributes for predicting question popularity. In fact, they strongly correlate with the topics. For example, fraction of questions starting with "Why" ranges from 1.8 to 20.5%, "How" – from 6.5 to 21.4%, and "I" – from 16.8 to 34.3%. We will address this below in more detail.

## PREDICTING POPULARITY
### Model Attribute Selection

One of declared goals of this study is to identify question attributes that can be edited by the AnswerXchange users in order to adjust the question popularity when needed. We start by arranging the model attributes (Table 1) into three groups. Group I includes the attributes that are responsible for semantic meaning of the question, time of the season and user type. By default, these attributes cannot be changed by the question asker. For example, a user who is asking a question about tax deduction is unlikely to change its topic from 10 to 18 (Figure 4) despite the fact that this change would increase the question popularity. Similarly, the user who is asking a question in April is unlikely to wait until next year when the question, if asked, may result in more popular question. Next, included to Group II are question attributes that can be changed as part of the question-asking experience. For example, regular user may adjust the length of question summary and/or details reshuffle the sentences when prompted. Adding proper capitalization and question mark is the standard practice employed by Q&A sites like Quora. Topic entropy can be changed, for example, by increasing the question length (Figure 6B). Although re-phrasing the question appears to be a more complicated task, regular AnswerXchange users can achieve success for some question types. For example, in a recent experiment the users were able to change "why" to another question type including even "closed-ended" questions [17]. Finally, included to Group III (not shown in Table 3) are text of the summary and text of details.

We applied Boruta algorithm [11] to 10% random sample of the 2015 AnswerXchange dataset to find all relevant features for Groups I and II (Table 3). Specifically, the algorithm produces pseudo-features (or "shadow attributes") from the existing features by shuffling the values of those features between the original samples. After generating the shadow attributes the procedure proceeds with building random-forest trees and comparing the Z-scores for the original features and shadow attributes. Based on this comparison, the algorithm decides whether a feature is confirmed important, tentatively important or unimportant. In Table 3, all but the last two attributes were confirmed important. Given the results shown in Figure 3B, the highest mean Z-score value is hardly surprising – a question asked earlier has far greater exposure to the search traffic compared to a question asked at the end of the tax season. While question topic feature was confirmed important, a relatively low mean Z-score is somewhat surprising. Three most important attributes from Group II are related to logarithm of text length (logarithm of text length plus 1 for the log of details length).

| Attribute | Type | Mean Z | Group |
|---|---|---|---|
| Week of the year | Categorical | 40.4 | I |
| Log of question length | Numeric | 30.3 | II |
| Log of details length | Numeric | 25.7 | II |
| Log of summary length | Numeric | 24.0 | II |
| First word of details | Categorical | 40.1 | II |
| First word of summary | Categorical | 15.3 | II |
| Coherency metrics | Numeric | 20.0 | II |
| Details flag | Binary | 14.6 | II |
| Platform | Categorical | 13.1 | I |
| Proper capitalization | Binary | 4.0 | II |
| Question mark | Binary | 5.1 | II |
| Question topic | Categorical | 3.4 | I |
| Product version | Categorical | 11.8 | I |
| Excessive capitalization | Binary | 4.8 | II |

Table 3. Popularity model attributes.

**Predictive Model**
When predicting the absolute number of views, it is helpful to switch from the regression problem (i.e. predicting the number of views or ranking of the question by popularity) to the classification problem (i.e. predicting if the question belongs to the top 1%, 5% or 10% of the most popular questions). In what follows, instead of predicting views, we are switching to a categorical judgment using a 10% cut-off value. Logistic regression is a form of regression in which one predicts the probability that an item belongs in one of two categories. Since probabilities are always between 0 and 1, one can threshold them on an empirically established number to produce class predictions following an approach similar to that used when predicting asker votes [15]. In such a way, the logistic regression scores do rank questions by the expected number of views. The ranks can be easily re-scaled to the absolute numbers of views if needed making the final results insensitive to the original categorization of popular vs. unpopular questions.

Model selection for this study is largely pre-defined by the Amazon Web Services (AWS), the cloud computing provider for TurboTax and AnswerXchange. As of 2015, the only option for binary classification offered by Amazon Machine Learning (AML) services was logistic regression. The AML services include variety of data transformations for the numeric and text model attributes; regularization; model testing and real time integration with Amazon Simple Storage Service (S3), MySQL and Amazon Redshift. By default, all numeric attributes are binned. Real time predictions are provided by AWS through "Enable for Real-time Prediction" interface.

**Model Performance**
Shown in Figure 7 are receiver operating characteristic (ROC) curves computed using the conventional validation with the 30% holdout dataset. The area under the curve (AUC) values for the ROC curves shown in Figure 7 are 0.678 (Group I), 0.759 (Groups I-II), and 0.793 (Groups I-III). Note that adding Group II attributes (black line in Figure 7) results in noticeable improvement in predictability of question popularity.

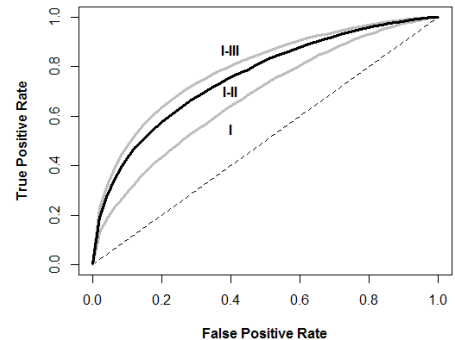

Figure 7. ROC curves of the AnswerXchange popularity models built with Group I (gray line), Group I-II (black line) and Group I-III (gray line) attributes.

We conclude this section by comparing model attributes important for predicting content popularity and quality as measured by the proxy metrics of question askers' votes. The complete list of the askers' vote model attributes has been presented in [15]. The most important attributes for the popularity model are the question details, and first word of the summary and summary length. The model attributes that are important for predicting the askers' vote are typically irrelevant for predicting popularity and vice versa. The only exception from this rule is the question summary length. Longer summaries typically result in better answers [15] that get voted up more frequently by both askers and viewers. Shorter summaries, on the other hand, are more popular but, understandably, may result in the answers that are less satisfying for the AnswerXchange users due to potentially missing context. The Pearson's correlation between askers' votes (1 for the up votes and 0 otherwise) and question popularity (1 for the questions in the top 10% and 0 otherwise) is a mere 0.065.

One way to explain this is to note that judging the answer quality may involve more human reasoning and cognitive efforts compared to selecting a particular question to click. Askers may agree or disagree with usefulness or perceived quality of the answer, but the relationship between details and search result ranking is controlled the search engine. Unless the users examine search results all the way to the bottom of the page (that is unlikely for a typical user), they may never find a question without details if a semantically similar question with details is already shown at the top.

## APPLICATIONS OF THE MODEL
### Trusted Users' Experience

As mentioned earlier, the majority of good quality answers in AnswerXchange come from trusted users (i.e. super users and agents). Support agents are salaried Intuit employees are therefore expected to work on any content creation or management task assigned to them. Super users are volunteers who only participate in the tasks of their own choosing, but they were demonstrated previously to be willing to engage in the content classification tasks. For example, super user's engagement to the content type classification (i.e. adding "tax" and "product" class labels to the AnswerXchange questions) was 50% [15].

Application of content popularity model to the task of improving content popularity is straightforward. First, the AnswerXchange questions would be divided into three segments: (1) content of high popularity no requiring any further actions, (2) content of low popularity that can be improved by modifying question writing style, and (3) content of low popularity that cannot be improved by modifying question writing style. The popularity model would be used in real time to compute content popularity score and flag the questions that fall into segment 2, i.e. the questions requiring re-phrasing by the trusted users as part of the question-answering experience or in a separate application. The model can be run as many times as needed to guide trusted users into generating popular content.

To illustrate this point, let us use a constructed example based on the real AnswerXchange question with summary "i need to add my health insurance info and i already sent my taxes through turbo tax. can i just add the medical info and resend my taxes through turbo tax" and without details. The question was asked during week 11, belongs to the most popular topic 18 (Figure 4B), does not have a question mark and not capitalized. The logistic regression score of this question is 0.016 and minor edits (proper capitalization and adding question mark) barely change it. Placing the second sentence ("Can I just add the medical info and resend my taxes through turbo tax?") ahead of the first one in the question summary increases the score to 0.073. Finally, moving the first sentence ("I need to add my health insurance info and I already sent my taxes through turbo tax.") to details and keeping the second sentence in the summary, increases the logistic regression score to 0.127.

### Regular Users' Experience

Another option to crowd source the task of question editing is to leverage question-asking experience. Users who ask questions in AnswerXchange are typically working on the tax returns and need an answer as quickly as possible. It is highly unlikely they will engage in the task of improving the AnswerXchange question popularity in the same manner as the trusted users. Can regular users be guided into creating popular content implicitly by personalizing question-asking experience? According to the results presented earlier, adding details to question typically results in increased question popularity and the 2015 AnswerXchange dataset provides additional insights into question writing behavior. Let us consider the users who did not notice "Add more details" option (Figure 1) as the control group and the users who did notice his option and used it as the test group. We now can apply uplift modeling framework to identify the user segments behind the most popular questions, i.e. the users who employed "Add more details" option in the most efficient way.

Uplift (or "incremental response") modeling [21] is a data mining technique used to predict not outright behavior, but rather the likelihood that an individual behavior will be influenced by an intervention or treatment. In recent years, uplift modeling has been applied in the financial services, telecommunications and retail industries to up-sell, cross-sell, churn and retention activities [21]. In a more recent example, uplift modeling went beyond traditional marketing tasks by identifying persuadable swing voters during President Barak Obama's 2012 campaign [21]. The theoretical foundations of uplift modeling and related algorithms have been reviewed by [7, 19, 21].

As mentioned earlier, the purpose of uplift modeling is to predict the reaction of the user to a treatment typically modeled by a binary variable. The treatment is expected to be a random number independent of the model attributes [19]. Additionally, treatment should not impact model attributes (in other words, model attributes should not change in response of the treatment) [19]. The latter requirement can be satisfied by adding constraint of the question length, i.e. by excluding summary length and details length from the list of model attributes. The equal split between "test" and "control" is not a requirement, although it maximizes statistical significance of the results.

Adding details to short questions aligns well with the uplift modeling paradigm as it is modeled by a binary variable and the user experience modifications would be relatively easy to implement. In fact, this would be equivalent to advising the user (who invested time in writing a long summary) to split the summary into sentences and move one or two sentences from the summary to details. Note that adding details in this case would not change the overall length of the question. When question length is limited by 170 characters (Figure 1) the response of the model may be mixed. Splitting long summaries would result in more views (Figure 6), while splitting short summary might work in the opposite direction. The goal of uplift modeling would be therefore to reveal the user and question features when intervention (i.e. splitting the summary) would make sense. By selecting only the first question asked by a user in 2015, we apply the model to the users rather than to questions to satisfy formal uplift modeling requirements. This reduces the 2015 AnswerXchange dataset to 160,500 questions, out of which 20.5% have details.

The implementation of the uplift modeling is based on the causal conditional trees [19] that estimate treatment effects (or uplift) by binary recursive partitioning in a conditional inference framework. Numerical example reported in this study has been generated with "uplift random forests" from the R package "uplift" [8]. The model outputs presented below are the scores and performance metrics (typically five or ten groups of questions comparable in size ranked by the gains in descending order). The calculated gains are used to plot the incremental gains curves [19].

The results of uplift modeling are shown in Figure 8. The left panel (Figure 8A) is the histogram of predicted uplift scores (the difference between fractions of questions in top 10% between questions with and without question details). The histogram exhibits a well-defined maximum close to 0.1 implying that treatment typically results in gains for the majority of the questions. Figure 8B shows the incremental gains curve [19]. The interpretation of the incremental gains is straightforward. If the curve follows the diagonal line, the treatment has the same effect across the sample. If, on the other hand, the incremental gains go up faster than diagonal line, certain segments are more receptive to the treatment. When incremental gains curve is flat or goes down, this is an indication of "anti-persuadable" behavior – the users who add details to questions belonging to the related segments are more likely to reduce number of views.

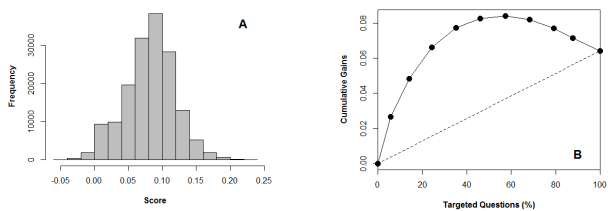

**Figure 8. Uplift scores (A) and incremental gains curve (B) for the uplift model.**

Admittedly, there are some limitations imposed by the fact that treatment is not random and may be correlated with other aspects of question writing. On the other hand, we are not planning to make the uplift model presented in this study to be operational in the traditional marketing sense. Rather, we are exploring new approaches to optimize the AnswerXchange question and answer experiences.

| Attribute | Importance (%) |
|---|---|
| Topic | 31.1 |
| Question length | 23.6 |
| Week number | 21.5 |
| First word of summary | 20.1 |
| Summary capitalization | 2.4 |
| Summary punctuation | 1.3 |

**Table 4. Importance of uplift model attributes.**

The results shown in Figure 8 and Table 4, demonstrate that redistributing the question between summary and details text fields keeping the question length unchanged results in an increased popularity of the question in 60% of cases. This approach is most effective when the summary length is close to 170 characters limit (as evidenced by the steep initial increase of the incremental gains curve). A the contrary, when summary length is relatively short, adding details to the question at the expense of shortening the summary results in just an opposite – the question popularity with the re-phrased question would be lower than for the question with the original one. An action to prevent this from happening would be to allow the regular user to post question as is. Capitalization and punctuation barely affect the uplift model predictions (although they still remain important attributes for the predictive model for content popularity).

## DISCUSSION AND CONCLUSION

Social Q&A systems offer numerous advantages for customer support operations including a substantial self-help component. For example, 99% of AnswerXchange users can find an answer to their particular question by simply searching for the asked similar questions. On the other hand, the content moderation in Q&A systems is largely performed reactively after the answer data has been generated and sufficient vote statistics and other feedbacks have been accumulated. As a result, poorly rated or low quality content is often removed only after it has potentially been viewed by a large number of searching users. Furthermore, the existing methods for identifying poorly rated content are often based on the assumption that the user feedback is objective and logical and not based on emotion or frustration. This is often not the case. While users' votes in AnswerXchange are predicable based on the attributes of the question [17], this is to a large degree due to an artifact – the users up vote tax related content more often than product related one therefore confusing quality of AnswerXchange answers and TurboTax experience.

Users' vote statistics incorporated in the AnswerXchange Lucene ranking algorithm has a limited impact on content popularity. Recall that AnswerXchange topics with high up vote fractions are not necessarily the most popular and vice versa (Figure 4). What is more, question features correlated with higher users' satisfaction may result in decreased popularity. Longer summaries typically results in better answers and hence more frequent up votes from askers', but tend to decrease popularity of AnswerXchange questions. Further research is needed to understand both rational and irrational factors behind AnswerXchange views and votes in order to provide the means of intervening into content creation process and altering the biased user behavior. Predictive modeling provides a convenient option to select attributes and rank them by importance. This was the approach employed in this study where we developed content popularity model complimenting out earlier effort in predicting asker votes in AnswerXchange [15, 16].

Combining predictive models provides extra opportunities for AnswerXchange operations. In fact, predicting both AnswerXchange question popularity and the askers' votes is equivalent to predicting the absolute numbers of up and down viewers' votes. In other words, if the number of question views over a pre-defined period of time is predictable, this would provide an estimate for the numbers of up and down votes over the same period of time. One application of this approach is the ability to interactively adjust the popularity of the question as part of the question formation process [17]: the user submitting a low quality question could be asked to modify it in such a way that its expected popularity would go down, whereas the user crafting a good quality question could be asked to re-phrase the question so that it will get more views. The benefit of such an approach is that not only can the individual asker's satisfaction with an answer be predicted and pro-actively addressed before the answer is even provided, but the satisfaction of other searching users with can be predicted as well to ensure that answers with expected poor satisfaction ratings are not provided to the users.

The main advantage of the uplift modeling approach is the ability to identify "persuadable" and "anti-persuadable" users [21] and so uplift scores may provide question-asking experience personalization in real time in order to increase the likelihood of the question to become popular. One option is to apply uplift modeling to all questions including the long-tail ones, i.e. irrespective to the question popularity. Another option is to improve content quality as well, but giving more weight to the popular content. The numbers of views per question may vary by orders of magnitude resulting in the same inequality in votes per question. Consequently, an unpopular question of low quality may be less damaging to the AnswerXchange operations as compared to a popular question of mediocre quality. Thus, predicting the expected question popularity and adjusting it as part of the question-asking experience may mitigate the impact of the low quality content. Recall that removing question details always results in decreased popularity and vice versa. Finally, one can also consider slightly more intrusive interventions to the question-asking experience. For example, instead of removing the question details, the user may be asked to re-phrase the first few words of the question in order to decrease its popularity.

The applicability of uplift modeling depends on several assumptions about the treatment and uplift model attributes. While the treatment is expected to be random, this requirement was not strictly enforced in this study. Based on the results of uplift modeling, however, we do see that askers' response to the treatment is far from uniform implying that agents and super users are not perfect in deciding on the optimal answer for a given question. This provides the rationale for our findings to be employed for AnswerXchange operations. The first option is to apply uplift modeling for generating answer tips in real time. For example, AnswerXchange can recommend including or excluding a web link or other type of reference to the answer based on the semantic analysis of the question. The second (more intrusive) option is to run A/B tests with a relatively small subset of answers and by assigning treatment variable at random. Since it would be difficult to apply full-scale A/B testing without major disruptions of the AnswerXchange user experience, the uplift modeling can be used to identify the smaller size test segments of the content with the largest gains in either direction.

There are several remaining tasks related to better understanding the mechanisms driving popularity of AnswerXchange questions. Recall that popularity model was designed to account for both (1) the degree to which the user pays attention to the question, intentionally or by mistake, and (2) the search engine placement of the question. The remaining research question to address is what are the relative impacts of Lucene ranking [14] and snippet readability features (e.g. the first token and length of the summary) on the user decision to click on the AnswerXchange search results? The same question applies to AnswerXchange content shown by Google and other search engines. The summary features may also influence the agent and super user decision to select a question for answering. Being answered is obviously a pre-requisite for the AnswerXchange question to become popular. We leave the remaining analysis for follow-up study.

The results reported in this paper are specific to the AnswerXchange. Other Q&A systems are designed differently and so question summary and details may be less important features with respect to question popularity. Our next task, however, will be to extend the findings reported in this paper to other AnswerXchange and TurboTax operations. For example, AnswerXchange askers may react differently to assertive answers and web references in answers [15] and so answer personalization based on uplift modeling can be considered. In addition to predicting expected askers' votes and question popularity, one can predict the likelihood of the assisted support contact or the likelihood of leaving positive feedback for TurboTax. The uplift modeling task would be to modify the TurboTax experience for the users who can be "persuaded" to search instead of contacting assisted support or to leave positive instead of negative product feedback.